\documentclass[twocolumn,separability,Amsterdam, nobibnotes, aps, pra,showpacs,preprintnumbers]{revtex4-1}
\usepackage{amssymb}
\usepackage{array}
\usepackage{graphicx}

\def\beq{\begin{equation}}
\def\eeq{\end{equation}}
\def\bea{\begin{eqnarray}}
\def\eea{\end{eqnarray}}

\begin{document}

\title{ Generating five dimensional Myers-Perry black hole solution using Quaternions}%

\author{Zahra Mirzaiyan}
\email{Z.Mirzaiyan@ph.iut.ac.ir}
\author{Behrouz Mirza}
\email{b.mirza@cc.iut.ac.ir}
\author{Elham sharifian}
\email{e.sharifian@ph.iut.ac.ir}

\affiliation{Department of Physics,
Isfahan University of Technology, Isfahan 84156-83111, Iran}


\begin{abstract}
Newman-Janis and Giampieri algorithms are two simple methods to generate stationary rotating black hole solutions in four dimensions. In this paper, we obtain the Myers-Perry black hole from the Schwarzschild solution in five dimensions by using quaternions. Our method generates the Myers-Perry black hole solution with two angular momenta in one fell swoop.\\
\end{abstract}

\maketitle

\section{Introduction}

Among all the different solutions of Einstein's equations, finding rotating ones has always been very challenging. In 1965, Newman and Janis (NJ) \cite{NJ} proposed a simple algorithm to obtain
axisymmetric rotating black hole metrics from a (spherical) static one through a particular complexification of radial and time coordinates, followed by a complex coordinate transformation. It seems that complexification is mostly based on intuitive conjectures as there is as of yet no reason claimed why this method works. The discovery of Kerr-Newman metric from a Reissner-Nordstrom solution (charged black hole) was the first achievement of NJ algorithm \cite{KN}. This algorithm has been used to generate different rotating solutions. More explanations and detailed reviews can be found in \cite{tal,Erbin,Erbin2,Erbin3,Erbin4,Whisker, Schiff,Fer}. In 1999, Kim was able to derive the $AdS_3$ rotating solution from the non-rotating one of Banados, Teitelboim, and Zanelli (BTZ) \cite{Kim} by applying the Newman-Janis formulation. Moreover, lots of efforts have been devoted to the generalization of rotating solutions to higher dimensions \cite{emparan,adam}.\\

Also, Krori and Bhattacharjee \cite{Krori} have shown that the NJ algorithm could be applied to the Brans-Dicke theory.\\

The NJ method can also be used to find the interior solutions of black holes. However, this method is used in \cite{Turrolla} to obtain some information about the interior solutions of Kerr metric.\\

The NJ algorithm is in fact not easy to perform because it requires the initial metric to be inverted in order to find out a set of null tetrad bases which implies implementing complex coordinate transformations. Finding out such a set of tetrad bases is rather tedious. An alternative useful algorithm has been proposed by Giampieri \cite{Giampieri} which requires neither the metrics to be inverted nor a null
tetrad basis to be found in order to apply the complex change of coordinates. Giampieri's method is especially useful when one has to work in arbitrary dimensions. Many attempts have been made to generalize solutions to higher dimensions for metrics with one angular momentum \cite{Xu}.\\

Recently, Erbin and Heurtier \cite{Erbin} presented a prescription for five dimensions with two angular momenta. They used Giampieri's simplification successively to obtain a five-dimensional Myers-Perry solution \cite{Myersperry} out of the Schwartzchild static one. In this process, they first rotated the $({\phi},r)$ plane while the $({\psi},r)$ one was kept fixed, then fixed the first plane and rotated the $({\psi},r)$ plane. In this way, they achieved a five-dimensional Myers-Perry solution via a two-step calculation.\\

In this paper, we propose a new method for finding rotating solutions from static ones in one fell swoop using quaternions. Our method is much simpler to use and more straightforward than that proposed in \cite{Erbin} for generating a five dimensional Myers-Perry black hole solution.
An interesting aspect of this novel method is its use of quaternions for the first time for generating a rotating solution with two angular momenta. Quaternions have a noncommutative algebra and we use symmetric products to remove the ambiguities in algebraic expressions.\\

This paper is organized as follows. In Section II, we review NJ and Giampieri methods. Quaternions and their mathematical features are introduced in Section III. Our method for generating the Myers-Perry black hole solution is discussed in Section IV and, finally, Section V is devoted to conclusions.

\section{Newman-Janis algorithm }

In this section, we review the Newman-Janis approach to the problem of deriving the
Kerr metric from the Schwarzschild solution based on the original work in 1965 \cite{NJ}. This algorithm presents an alternative way for deriving rotating solutions by applying a particular complex transformation to a non-rotating static metric in four dimensions. More details may be found in \cite{Erbin2}. To use this algorithm, we consider the usual form of a static metric in coordinate $(t, r, \theta, \phi)$ as follows:
\begin{equation}
ds^{2}=-f(r) dt^2 +\frac{dr^{2}}{f(r)}+r^2 d{\Omega}^{2},
\end{equation}
where,
\begin{equation}
\ \ d{\Omega}^{2}=d{\theta}^{2}+{\sin^{2}{\theta}}\ \ d{\phi}^{2}.
\end{equation}
Applying the transformation to the Edington-Finklestein retarded null coordinates with $u=t-r_{*}$ defined as $r_{*}=\int{dr (\frac{-g_{rr}}{g_{tt}})^{\frac{1}{2}}}$ or equivalently as:
\begin{equation}\label{du}
du=dt- f(r)^{-1} dr,
\end{equation}
\noindent the metric takes the following form:
\begin{equation}\label{newmetric}
ds^{2}=-f(r) du^{2}-2 du dr +r^{2} d{\Omega}^{2}.
\end{equation}
We next introduce the set of null tetrads $l_{\mu},n_{\mu},m_{\mu},\bar{m}_{\mu}$ satisfying the following conditions
\begin{equation}\label{tetradcondition}
l_{\mu} n^{\mu}=-m_{\mu} {\bar{m}}^{\mu}=-1,
\end{equation}
while other scalar products vanish.
The metric can be rewritten in terms of null tetrads as $g^{\mu\nu}=l^{\mu} n^{\nu}+n^{\mu} l^{\nu}-m^{\mu} {\bar{m}}^{\nu}-{\bar{m}}^{\mu} m^{\nu}$. For the metric in (\ref{newmetric}), the tetrades are determined as:
\begin{eqnarray}\label{tetrad}
l^{\mu}&=&{\delta}_{r}^{\mu},\nonumber\\
n^{\mu}&=&{\delta}_{u}^{\mu}-\frac{f(r)}{2}{\delta}_{r}^{\mu},\nonumber\\
m^{\mu}&=&\frac{1}{\sqrt{2}\ {\bar{r}}}({\delta}_{\theta}^{\mu}+{\frac{i}{\sin{\theta}}}{{\delta}_{\phi}^{\mu}}),\nonumber\\
{\bar{m}}^{\mu}&=&\frac{1}{\sqrt{2}\ {r}}({\delta}_{\theta}^{\mu}-{\frac{i}{\sin{\theta}}}{{\delta}_{\phi}^{\mu}}),
\end{eqnarray}\\

\noindent where, ${\bar{m}}^{\mu}$ is the complex conjugate of $m^{\mu}$ and $r\in R$ such that $r=\bar{r}$.
Now, the components $u$ and $r$ are allowed to take complex values simultaneously subject to the following conditions:\\

\noindent 1. During the transformation, $l^{\mu}$ and $n^{\mu}$ are kept real;\\
\noindent 2. $m^{\mu}$ and ${\bar{m}}^{\mu}$ must be still complex conjugates to each other; and\\
\noindent 3. One could recover the previous basis for $r\in R$.\\

We now perform the process of ``complex conjugate transformation'' for the components $u$ and $r$ as follows:

\begin{eqnarray}
u&=&u^{\prime}+ia\ \cos{\theta},\nonumber\\
r&=&r^{\prime}-ia\ \cos{\theta},\nonumber\\
{\theta}&=&{\theta}^{\prime}, \nonumber\\
{\phi}&=&{\phi}^{\prime}.
\end{eqnarray}

Here, $a$ is a real parameter and $u^{\prime}$ and $r^{\prime}$ are real too. Also the form of the function $f(r)$ in the initial metric is invariant under this transformation.\\
Under this transformation, the new tetrads take the following form
\begin{eqnarray}\label{newtetrad}
l^{{\prime}{\mu}}&=& {\delta}_{r}^{\mu},\nonumber\\
n^{{\prime}{\mu}}&=&{\delta}_{u}^{\mu}-\frac{f(r)}{2}{\delta}_{r}^{\mu},\nonumber\\
m^{{\prime}{\mu}}&=&\frac{1}{\sqrt{2} {(r+ia\ cos{\theta})}}({\delta}_{\theta}^{\mu}+{\frac{i}{\sin{\theta}}}{{\delta}_{\phi}^{\mu}}-ia\ sin{\theta}({\delta}_{u}^{\mu}-{\delta}_{r}^{\mu})),\nonumber\\
{\bar{m}}^{{\prime}{\mu}}&= &\frac{1}{\sqrt{2} {(r-ia\ cos{\theta})}}({\delta}_{\theta}^{\mu}-{\frac{i}{\sin{\theta}}}{{\delta}_{\phi}^{\mu}}+ia\ sin{\theta}({\delta}_{u}^{\mu}-{\delta}_{r}^{\mu})).\ \ \ \
\end{eqnarray}
The next step in the Newman-Janis algorithm is to construct the inverse metric $g^{{\mu}{\nu}}$ using the new tetrads (\ref{newtetrad}), the metric is converted to obtain the covariant components of the rotating solution from the initial static one.
Using the above method for the familiar Schwarzschild solution with $f(r)=1-\frac{2m}{r}$ in four dimensions, we obtain the Kerr metric as

\begin{eqnarray}\label{newker}
d{{s}^{2}}&=&(1-\frac{2mr}{{{r}^{2}}+{{a}^{2}}{{\cos }^{2}}\theta})d{{u}^{2}}+2dudr\nonumber\\
&& +\frac{4mra\ {{\sin }^{2}}\theta }{{{r}^{2}}+{{a}^{2}}{{\cos }^{2}}\theta }dud\phi -2a\ {{\sin }^{2}}\theta d\phi dr\nonumber\\
&&-(({{r}^{2}}+{{a}^{2}}{{\cos }^{2}}\theta ){{a}^{2}}{{\sin }^{2}}\theta +2mr{{a}^{2}}{{\sin }^{2}}\theta\nonumber\\
&&+{{({{r}^{2}}+{{a}^{2}}{{\cos }^{2}}\theta )}^{2}})\frac{{{\sin }^{2}}\theta }
{({{r}^{2}}+{{a}^{2}}{{\cos }^{2}}\theta )}d{{\phi }^{2}}\nonumber\\
&&-({{r}^{2}}+{{a}^{2}}{{\cos }^{2}}\theta )d{{\theta }^{2}}.
\end{eqnarray}

In 1990, G. Giampieri proposed a simplification to the Newman-Janis method \cite{Giampieri}. In Giampieri's version of the algorithm, one could avoid using null tetrads to work directly with the metric. Similar to the Newman-Janis algorithm, the null coordinate $u=t-r_{*}$ must be introduced before the metric is rewritten as follows:
\begin{equation}\label{metricinu}
ds^{2}=-f(r) du^{2}-2 du dr+r^2d{\Omega}^{2}.
\end{equation}
Now, the coordinates $u$ and $r$ are allowed to be complex and transformed as in (\ref{transformation}) below:
\begin{eqnarray}\label{transformation}
u&=&u^{\prime}+ia\ \cos{\psi},\nonumber\\
r&=&r^{\prime}-ia\ \cos{\psi},\nonumber\\
{\theta}&=&{\theta}^{\prime},\nonumber\\
{\phi}&=&{\phi}^{\prime}.
\end{eqnarray}
By introducing the new real angle ${\psi}$, we embed the four dimensional space-time into a 5-dimensional complex space-time. The differentials then read as follows:
\begin{eqnarray}
du= du^{\prime}-ia\ \sin{\psi}\ d{\psi},\nonumber\\
dr= dr^{\prime}+ia\ \sin{\psi} \ d{\psi}.
\end{eqnarray}
Substituting these coordinates in metric (\ref{metricinu}), we have a new metric of the following form
\begin{eqnarray}
ds^{2}&=&- \tilde{f}(r) (du-ia\ \sin{\psi}\ \ d{\psi})^{2}\nonumber\\
&&-2 (du-ia\ \sin{\psi}\ \ d{\psi})(dr+ia\ \sin{\psi} \ d{\psi})\nonumber\\
&&+ (r^{2} +a^{2} \cos^{2} {\theta}) d{\Omega}^{2}.
\end{eqnarray}

The above metric is not real. It may be reduced to a real desired result by applying the ansatz
\begin{equation}
i d{\psi}=\sin{\psi}\ d\phi,\ \ {\psi}={\theta}.
\end{equation}

Clearly, the last step is helpful and will help us in a neat manner to obtain the Kerr metric from a Schwarzschild one as follows:

\begin{eqnarray}\label{kerr}
ds^{2}&=&f(r) (du-a\ \sin^{2} {\theta}\ d\phi)^{2}\nonumber\\
&&-2(du-a\ \sin^{2} {\theta}\ d\phi)(dr+a\ \sin^{2} {\theta}\ d\phi)\nonumber\\
&&+ {{\rho }^{2}}d{\Omega}^{2},
\end{eqnarray}
where we have defined
\begin{equation}
{{\rho }^{2}}={{r}^{2}}+{{a}^{2}}{{\cos }^{2}}\theta.
\end{equation}
The final step is to redefine the standard coordinate and go to the Boyer-Linqiust coordinate using the following transformations

\begin{eqnarray}\label{goto boyer}
du&=&dt^{\prime}-g(r) dr,\nonumber\\
d{\phi}&=& d{\phi}^{\prime}-h(r) dr.
\end{eqnarray}

The conditions ${{g}_{tr}}={{g}_{r{\phi }'}}=0$ are applied for

\begin{equation}
g=\frac{r^2 +a^2}{\Delta},\ \ h=\frac{a}{\Delta},\nonumber\\
\end{equation}

where ${\Delta}$ is introduced as
\begin{equation}
{\Delta} = \tilde{f}(r) {\rho}^2+a^2 \sin^2 {\theta}.
\end{equation}
It is noted that functions $g(r)$ and $h(r)$ are dependent only on the coordinate $r$. Finally, by omitting the primes, we get the Kerr metric as
\begin{eqnarray}
d{{s}^{2}}&=&-\tilde{f}(r)d{{t}^{2}}+\frac{{{\rho }^{2}}}{\Delta }d{{r}^{2}}+{{\rho }^{2}}d{{\theta }^{2}}+\frac{{{\Sigma }^{2}}}{{{\rho }^{2}}}{{\sin }^{2}}\theta \ d{{\phi }^{2}}\nonumber\\
&&+2a(\tilde{f}-1)\ {{\sin }^{2}}\theta \ dtd\phi.
\end{eqnarray}

Recently, \cite{Erbin} proposed a way for obtaining the Myers-Perry solutions \cite{Myersperry} from the five dimensional Schwarzschild metric by applying the Giampieri's method in two iterations. \\ Our main purpose in this paper is using quternions to obtain the Myers-Perry solution in five dimensions. In \cite{Erbin}, the simplification is accomplished in the $(r,{\phi})$ plane while the $({\psi},r)$ plane is kept fixed followed by
fixing the first plane and doing the rotation in the $(r,{\psi})$ plane. A two-step calculation yields a five-dimensional Myers-Perry solution. Our new method does not require the successive use of Giampieri's approach so that the same using our simplification and quaternions with calculations in only one fell swoop.
The next section briefly introduces the quaternions.\\

\section{Quaternions}

Quaternions comprise number systems that extend complex numbers in mathematics. The systems may be perceived as usual orthogonal coordinates in which complex unit vectors are used in place of the unit vectors used in previous methods. Quaternions are generally represented in a specific form as follows:

\begin{equation}\label{quater}
a+b\ i +c\ j+d\ k,
\end{equation}
where, $a,b,c,$ and $d$ are real numbers and $i, j,$ and $k$ are quaternion units. This is exactly like writing a vector in terms of unit vectors. Fundamental unit quaternions have the following feature:

\begin{equation}
i^{2}=j^{2}=k^{2}=-1.
\end{equation}
Note that $i$,$j$ and $k$ are imaginary numbers which we can define the conjugate of $i$ as $i^{*}=-i$.\\
Also, the multiplication of two quaternions is not commutative, reminding us of the external multiplication laws. Quaternion multiplications are presented in Table 1.
\begin{table}
\caption{Quaternion multiplication}
\begin{center}
\begin{tabular}{ | m{2em} | m{2em}| m{2em}| m{2em}|m{2em}|}
\hline
${\times}$ & 1 & i & j & k \\
\hline
1 & 1 & i & j & k \\
\hline
i& i & -1 & k & -j \\

\hline
j & j & -k & -1 & i \\

\hline
k & k & j & -i & -1\\

\hline
\end{tabular}
\end{center}
\end{table}

The next Section introduces the proposed method that employs the noncommutative characteristic of quaternion multiplication to recover the Myers-Perry solution in five dimensions in one fell swoop.\\

\section{Derivation of Myers–-Perry solution in five dimensions }\label{S4}

In this Section, we recover the rotating Myers-Perry solution \cite{Myersperry} with two angular momenta in five dimensions using quaternions. This application of quaternions is new and help us to obtain these solutions in one fell swoop.
Let us start with a five dimensional Schwarzschild metric as follows:
\begin{equation}\label{5sch}
ds^{2}=-f(r) dt^{2}+{f(r)}^{-1} dr^2+r^{2} d{\Omega}^{2} _{3},
\end{equation}

\noindent where, $d{\Omega}^{2} _{3}$ is the metric induced on the sphere $S^{3}$ in the Hopf coordinate to be read as follows:

\begin{equation}\label{domeg}
d{\Omega}^{2} _{3}=d{\theta}^{2}+ \sin^{2} {\theta}\ d{\phi}^{2}+ \cos^{2} {\theta}\ d{\psi}^{2},
\end{equation}
and
\begin{equation}
f(r)=1-\frac{m}{r^{2}}.
\end{equation}

Its transformation to the Eddington-Finklestein retarded null coordinate leads us to the following metric:

\begin{equation}\label{newm}
ds^{2}=-du (du+2dr)+(1-f(r)) du^{2}+r^{2} d{\Omega}^{2} _{3}.
\end{equation}

The next step involves the complex coordinate transformation. In this way, quaternions are used to propose new transformations for the $u$ and $r$ coordinates.

\begin{eqnarray}\label{newtr}
u&=& u^{\prime}+i a\ \cos{{\chi}_{1}}+j b\ \sin {{\chi}_{2}},\nonumber\\
r&=& r^{\prime}-i a\ \cos{{\chi}_{1}}-j b\ \sin {{\chi}_{2}}.\nonumber\\
\end{eqnarray}
Based on the above ansatzs, we have:
\begin{eqnarray}
du&=&du^{\prime}-ia\ \sin{{\chi}_{1}}\ d{{\chi}_{1}} +j b\ \cos{{\chi}_{2}}\ d{{\chi}_{2},}\nonumber\\
dr&=&dr^{\prime}+ia\ \sin{{\chi}_{1}}\ d{{\chi}_{1}} - j b\ \cos{{\chi}_{2}}\ d{{\chi}_{2}.}\nonumber\\
\end{eqnarray}

In (\ref{newtr}), $i$ and $j$ are quaternions. Also, $a$ and $b$ are two parameters to be seen later that they are related to independent angular momenta. ${\chi}_{1}$ and ${\chi}_{2}$ are two angles introduced here solely to help us perform the simplification method. As we can see from (\ref{newtr}), the transformation takes place simultaneously in both $(r,{\phi})$ and $(r,{\psi})$ plane. They will vanish as a consequence of the slice fixing process and the following ansatz.\\
\begin{eqnarray}\label{GA}
i d{{\chi}_{1}}&=&\sin{{\chi}_{1}} \ d{\phi},\ \ {{\chi}_{1}}={\theta},\nonumber\\
j d{{\chi}_{2}}&=&-\cos{{\chi}_{2}} \ d{\psi},\ \ {{\chi}_{2}}={\theta}.
\end{eqnarray}

This transformation means the simultaneous transformation of both $(r,\phi)$ and $(r,\psi)$ planes.
Relation (\ref{newtr}) should be now replaced with the angle fixings in (\ref{GA}) in the metric in (\ref{newm}).\\

\noindent With the new sets of transformations (\ref{newtr}), we have $r^{2}=r r^{*}=r^{{\prime}{2}} +a^{2} \cos^{2} {\theta}+b^{2} \sin^{2} {\theta}$.

So, we have

\begin{eqnarray}\label{trametric}
ds^{2}&=&-(du^{\prime}-ia\ \sin{{\chi}_{1}}\ d{{\chi}_{1}} +j b\ \cos{{\chi}_{2}}\ d{{\chi}_{2}}) \times \nonumber\\
&& [((du^{\prime}-ia\ \sin{{\chi}_{1}}\ d{{\chi}_{1}} +j b \ \cos{{\chi}_{2}}\ d{{\chi}_{2}})+2(dr^{\prime}\nonumber\\
&&+ia\ \sin{{\chi}_{1}}\ d{{\chi}_{1}} - j b\ \cos{{\chi}_{2}}\ d{{\chi}_{2}}))]\nonumber\\
&& +(1-{\tilde{f}}(r^{\prime})) (du^{\prime}-ia\ \sin{{\chi}_{1}} d{{\chi}_{1}} +j b \ \cos{{\chi}_{2}}\ d{{\chi}_{2}})^{2}\nonumber\\
&& +\mbox{angular part of the metric}.
\end{eqnarray}

The functional form of $f(r)$ under quaternion complexification transformations will be as follows:
\begin{equation}\label{newf}
f(r^{\prime})=1-\frac{m}{r^{{\prime}{2}} +a^{2} \cos^{2} {\theta}+b^{2} \sin^{2} {\theta}}.
\end{equation}

\noindent The following three terms in the angular part need to be transformed.\\
\noindent 1.The term $r^{2} d{\theta}^{2}$
\begin{equation}\label{ft}
r^{2} d{\theta}^{2}\longrightarrow (r d\theta) ({r d\theta)^*} = (r^{{\prime}{2}} +a^{2} \cos^{2} {\theta}+b^{2} \sin^{2} {\theta})d{\theta}^{2}.
\end{equation}

\noindent 2. The second term is $r^{2} \sin^{2} {\theta} d{\phi}^{2}$.
Here, we have the angle $\phi$ and we know from the angle fixing conditions in (\ref{GA}) that the transformation in the $(r,\phi)$ plane should be accomplished via angle ${\chi}_{1}$
\begin{eqnarray}\label{st}
r^{2} \sin^{2} {\theta} d{\phi}^{2}\longrightarrow \sin^{2} {\theta} (r d{\phi}) (r d{\phi})^{\ast}.
\end{eqnarray}
We now proceed with the calculation of the term $r d {\phi}$ by replacing it with the following ansatz (see Eq.(\ref{GA})):
\begin{eqnarray}\label{dphi}
&&r d{\phi}= r (\frac{i d{{\chi}_{1}}}{\sin{{\chi}_{1}}})=(\frac{i\cdot r+r\cdot i}{2})( \frac{ d{{\chi}_{1}}}{\sin{{\chi}_{1}}})= \nonumber\\
&&[ (\frac{i\cdot (r^{\prime}-i a\ cos{{\chi}_{1}}-j b\ sin {{\chi}_{2}})+(r^{\prime}-i a\ cos{{\chi}_{1}}-j b\ sin {{\chi}_{2}})\cdot i}{2}) \nonumber\\
&&\times \frac{d{{\chi}_{1}}}{\sin{{\chi}_{1}}}] \nonumber\\
&&=\frac{i d{{\chi}_{1}}}{\sin{{\chi}_{1}}} ( r^{\prime} -i a\ cos{{\chi}_{1}}).
\end{eqnarray}
It should be noted that $i$ and $j$ are noncommutative and therefore we should use a symmetric form in all products to have well-defined relations.
The last line of the above relation was obtained by using the non-commutative feature of quaternion multiplication $(i \cdot j=-j \cdot i)$.\\
The transformed form of (\ref{st}) now reads as follows:
\begin{eqnarray}\label{st2}
&& r^{2} \sin^{2} {\theta} d{\phi}^{2}\longrightarrow \nonumber\\
&& \ \ \sin^{2} {\theta}\ \ ( r^{\prime} -i a\ \cos{{\chi}_{1}}) ( r^{\prime} + ia\ \cos{{\chi}_{1}}) \frac{ d{{\chi}_{1}^{2} }}{\sin^{2} {{\chi}_{1}}}\nonumber\\
&& =\sin^{2} {\theta} (r^{\prime} -i a\ \cos{{\chi}_{1}}) (r^{\prime} + i a\ \cos{{\chi}_{1}}) \ d{\phi}^{2}\nonumber\\
&&= \sin^{2} {\theta} (r^{{\prime}{2}} + a^2 \cos^{2} {\theta})\ d{{\phi}^{2}},
\end{eqnarray}
where, $d{\phi}^2=\frac{ d{{\chi}_{1}^{2} }}{\sin^{2} {{\chi}_{1}}}$ is used in the third line and ${\chi}_{1}=\theta$ in the last line which is the angle fixing condition in (\ref{GA}). The basic point in our method is the replacement of the products of noncommutative quaternions with a symmetric and well-defined form.\\

\noindent 3. The third term is $r^{2} \cos^{2} {\theta} d{\psi}^{2}$.
In this case, we have the angle $\psi$ and we know from the angle fixing conditions in (\ref{GA}) that the transformation in the $(r,\psi)$ plane needs to be accomplished via angle ${\chi}_{2}$.

\begin{eqnarray}\label{tt}
r^{2} \cos^{2} {\theta}\ d{\psi}^{2}\longrightarrow \sin^{2} {\theta}\ (r d{\psi}) (r d{\psi})^{\ast}.
\end{eqnarray}
The symmetrizing angle part method used with the angle fixing ansatz $d{\psi}=\frac{-j d{{\chi}_{2}}}{\cos{{\chi}_{2}}}$ helps us transform this term of the angular part of the metric as

\begin{eqnarray}\label{tt2}
r^{2} \cos^{2} {\theta}\ d{\psi}^{2}\longrightarrow \cos^{2} {\theta}\ (r^{{\prime}{2}} +b^{2} \sin^{2} {\theta})\ d{{\psi}^{2}}.
\end{eqnarray}

Using Eqs. (\ref{trametric}), (\ref{ft}), (\ref{st2}), and (\ref{tt2}), we can now obtain the transformed metric under the quaternions complexification process as follows (all the primes are omitted)
\begin{eqnarray}\label{lastmetric}
ds^{2}&=&-du^{2} -2 du dr \nonumber\\
&&+(1-\tilde{f} {(r)}) (du-a\ \sin^{2} {\theta}\ d{\phi} -b\ \cos^{2}{\theta}\ d{\psi} )^{2} \nonumber\\
&&+2a\ \sin^{2} {\theta}\ dr d{\phi}+2b\ \cos^{2} {\theta}\ dr d{\psi}+ {\rho}^{2} d{\theta}^{2}\nonumber\\
&&+(r^{2} +a^{2}) \sin^{2} {\theta}\ d{\phi}^{2}+(r^{2} +b^{2}) \cos^{2} {\theta}\ d{\psi}^{2}, \
\end{eqnarray}
where,
\begin{equation}\label{ro}
{\rho}^{2}=r^{2} +a^{2} \cos^{2} {\theta} +b^{2} \sin^{2} {\theta},
\end{equation}
and
\begin{equation}\label{f}
\tilde{f}(r)=1-\frac{m}{r^2 +a^{2} \cos^{2} {\theta} +b^{2} \sin^{2} {\theta}}.
\end{equation}

The metric can go to the Boyer-Lindquist coordinate through the following transformations:

\begin{eqnarray}\label{BLT}
du&=& dt-g(r) dr,\nonumber\\
d\phi &=& d{{\phi}^{\prime}}- h_{\phi}(r) dr,\nonumber\\
d\psi &=& d{{\psi}^{\prime}} -h_{\psi}(r) dr,\nonumber\\
\end{eqnarray}
where,
\begin{eqnarray}
g(r)&=&\frac{\Pi}{\Delta},\nonumber\\
h_{\phi}(r)&=&\frac{\Pi}{\Delta} \frac{a}{r^{2}+a^{2}},\nonumber\\
h_{\psi}(r)&=&\frac{\Pi}{\Delta} \frac{b}{r^{2}+b^{2}},
\end{eqnarray}
based on the following definition:
\begin{eqnarray}
\Pi &=&({r^{2}+a^{2}})({r^{2}+b^{2}}),\nonumber\\
\Delta &=&r^{4} + r^{2} (a^{2} +b^{2}-m)+a^{2} b^{2}.
\end{eqnarray}
Finally, we can find the final Myers-Perry metric in five dimensions as follows:
\begin{eqnarray}\label{fmetric}
ds^{2}&=&-dt^{2} +(1-\tilde{f}(r))(dt-a\ \sin^{2} {\theta}\ d{\phi} -b\ \cos^{2}{\theta}\ d{\psi} )^{2}
\nonumber\\
&&+{\rho}^2 d{\theta}^{2} +\frac{r^{2} {\rho}^{2}}{\Delta} dr^2+(r^{2} +a^{2}) \sin^{2} {\theta}\ d{\phi}^{2}+ \nonumber\\
&&+(r^{2}+b^{2}) \cos^{2} {\theta}\ d{\psi}^{2}.
\end{eqnarray}

In this way, we recover the Myers-Perry solution in five dimensions by using quaternions complexification transformations while we symmetrize the noncommutative products of quaternions. Clearly, application of quaternions in this approach helps us obtain rotating five-dimensional solutions in one fell swoop.

\section{Conclusion}\label{S6}
In this work, we proposed a new method for generating spinning black hole solutions based on quaternions' algebra. In this process, we used the main property of noncommutative product of quaternion. The Myers-Perry solution in the five-dimensional space-time was derived from the static Schwarzschild one.
The advantages of the proposed method over previous ones include: 1) it is easier to obtain
rotating solutions from static ones and 2) calculations can be performed in one fell swoop. It will be interesting to generalize this method to higher dimensions.


\end{document}